\documentclass{IEEEcsmag}

\usepackage[colorlinks,urlcolor=blue,linkcolor=blue,citecolor=blue]{hyperref}

\usepackage{upmath}
\usepackage{amsmath,amsthm,amsfonts,amssymb,amscd}
\usepackage[T1]{fontenc}
\usepackage{subcaption}

\usepackage{tcolorbox}%

\usepackage{xcolor}

\begin{document}

\editor{Editor: Name, xxxx@email}

\title{Datacenter Ethernet and RDMA: Issues at Hyperscale}

\author{Torsten Hoefler}
\affil{ETH Z\"urich and Microsoft}

\author{Duncan Roweth, Keith Underwood, Bob Alverson}
\affil{Hewlett Packard Enterprise}

\author{Mark Griswold, Vahid Tabatabaee, Mohan Kalkunte, Surendra Anubolu}
\affil{Broadcom}

\author{Siyuan Shen}
\affil{ETH Z\"urich}

\author{Moray McLaren}
\affil{Google}

\author{Abdul Kabbani, Steve Scott}
\affil{Microsoft}

\markboth{IEEE Computer}{Datacenter Ethernet and RDMA: Issues at Hyperscale}

\definecolor{mycolor}{rgb}{0.122, 0.435, 0.198}%
\makeatletter
\newcommand{\mybox}[1]{%
	\setbox0=\hbox{#1}%
	\setlength{\@tempdima}{\columnwidth}%
	\begin{tcolorbox}[colframe=mycolor,boxrule=0.5pt,arc=4pt,left=6pt,right=6pt,top=6pt,bottom=6pt,boxsep=0pt,width=\@tempdima]
		#1
	\end{tcolorbox}
}
\makeatother

\newcommand{\outlook}[1]{\mybox{\includegraphics[width=1.5em]{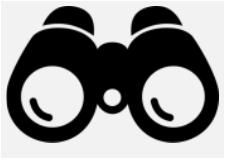} #1}}
\newcommand{\hl}[1]{{#1}}

\begin{abstract}
We observe that emerging artificial intelligence, high-performance computing, and storage workloads pose new challenges for large-scale datacenter networking.
RDMA over Converged Ethernet (RoCE) was an attempt to adopt modern Remote Direct Memory Access (RDMA) features into existing Ethernet installations.
Now, a decade later, we revisit RoCE's design points and conclude that several of its shortcomings must be addressed to fulfill the demands of hyperscale datacenters.
We predict that both the datacenter and high-performance computing markets will converge and adopt modernized Ethernet-based high-performance networking solutions that will replace TCP and RoCE within a decade. 
\end{abstract}

\maketitle

\newcommand{\para}[1]{\textcolor{blue}{[#1]}}
\renewcommand{\para}[1]{}

\section{Datacenter Ethernet's new Environment} 

Ethernet has dominated the wired local-area networking (LAN) space for decades ranging from deployments in private homes to the largest datacenters. Datacenters have experienced a massive growth during the last decade and the number of connected machines exceeds the size of the largest supercomputers today. While there remain some differences, the networking requirements of such hyperscale mega-datacenters and supercomputers are quite similar~\cite{mdchpc}. Yet, supercomputers are traditionally connected using special-purpose interconnects while datacenters build on Ethernet. Due to similar requirements and economies of scale, both continue to grow closer together with each new technology generation. We believe now is the right time to re-think the basic assumptions and architecture for a converged interconnect. 

Multiple technological trends are accelerating this convergence of high-performance interconnects. Primarily, the increasing network performance requirements push towards more efficient host stacks that can support the terabit bandwidths, hundreds of millions of transactions per second, and single-digit microsecond latencies that are required by emerging data-intensive applications such as Artificial Intelligence (AI)~\cite{hxmesh}. These extreme requirements force all protocols and hardware to be as \hl{efficient} as possible, ruling out many of the TCP/IP-like stacks that traditionally drove datacenter networking. Remote Direct Memory Access (RDMA) was developed nearly three decades ago for high-performance computing (HPC) workloads and was later expanded to target storage with InfiniBand (IB) Verbs RDMA.  RDMA enables CPU-offloaded, hardware-accelerated direct memory access over the network. During the last 10 years, it became the de-facto standard for low-overhead and high-speed networking. Nearly all supercomputer architectures as well as leading datacenter providers utilize RDMA in production today.

The simple assumptions on load balancing, congestion control, and error handling made decades ago do not hold for today’s networks that have more than 100x higher bandwidth and 10x higher message rates. Furthermore, simple RDMA network interface cards (NICs) are often enhanced with additional functionalities. The resulting “Smart NICs” often offload significant services and implement specialized network protocols. Modern network switches also have improved capabilities ranging from advanced in-network telemetry, in-network computation capabilities, and in-network load-balancing or congestion-control~\cite{sspaper}. We argue that the currently existing standards and deployed infrastructure has fundamental gaps that must be addressed in the near future to support efficient high-performance networking.

\section{A brief history of RDMA for Ethernet}

\hl{RDMA was originally developed for HPC in systems as early as the Paragon, Cray’s T3D/T3E, and ASCI Red. Later, InfiniBand Verbs RDMA became wide-spread in the supercomputing field as a standardized solution.} It was then adopted as “RDMA over Converged Ethernet” (RoCE) in the datacenter context to provide RDMA’s benefits in a backwards-compatible Ethernet context. Another protocol, iWARP (cf.~IETF 2007, RFCs 5040-5044, 6580, 6581, 7306), layers RDMA semantics over TCP or SCTCP. Both iWARP and RoCE use InfiniBand’s Verbs to interface with the user software stacks and are thus mostly transparent to the user. Even though iWARP allowed Internet-compatible routing from the beginning, it did not find widespread adoption. This may be due to the fact that a full TCP/IP stack is complex and expensive to offload to hardware, compared to the very simple protocol that underlies RoCE. Indeed, RoCEv1 was simply adopting an InfiniBand-like transport layer (i.e., the Base Transport Header, BTH) on top of Ethernet’s L2 headers. Later, RoCEv2 added IP/UDP L3 headers to support routing within and across datacenters. \hl{Today, there are more RoCEv2 NICs than InfiniBand NICs deployed.}

\subsection{RoCE – convergence or duct tape?}

\hl{RoCE’s core design is inherited from a technology developed for simple hardware two decades ago and are suboptimal in today’s Ethernet environments.} For example, RoCE uses InfiniBand’s simple transport layer that heavily builds on in-order delivery as well as go-back-n retransmission semantics that essentially require a highly reliable in-order fabric for efficient operation. Thus, RoCE runs best over a lossless in-order fabric, like InfiniBand. Traditionally, Ethernet drops packets when switch buffers are full and relies on end-to-end retransmission. To support RoCE, “converged Ethernet” (CE) introduces Priority Flow Control (PFC) to implement link-level lossless operation. PFC repurposes Ethernet PAUSE frames that existed in Ethernet to support networks with different link transmission rates. PFC enhances PAUSE frames to stop (or throttle) traffic on a specific priority class to avoid packet drops. %
Unfortunately, this complex set of protocols interferes across the different layers in the network and reduces efficiency for some of today’s most important workloads. 

RoCE’s semantics, load balancing, and congestion control mechanisms are inherited from InfiniBand. This implies that all messages should appear at the destination in order as if they were transmitted over a static route, essentially disallowing many packet-level load balancing mechanisms. For AI training workloads which are long-lived flows, \hl{multi-pathing mechanisms can greatly improve the job completion time}. Furthermore, RoCEv2 uses a simplistic congestion control mechanism based on IP’s Explicit Congestion Notification (ECN). ECN-compatible switches mark packets when congestion is detected and receivers relay that information back to the senders, which in turn reduce their injection rate guided with a single parameter. After a congestion-free period, the rate is automatically increased again using a second configuration parameter. ECN uses a binary flag for congestion experienced and the lack of fine grained indication results in many \hl{Round Trip Times (RTTs)} to determine the correct rate. This simple mechanism is very similar to InfiniBand’s original Forward and Backward Explicit Congestion Notification (FECN/BECN). It promises to coexist with other traffic but is hard to configure in practice~\cite{hpcc,dcqcn,timely}.

We now briefly discuss some important traffic motifs in HPC and datacenter traffic and then discuss RoCE’s shortcomings in detail. 

\section{Guiding Traffic Motifs}

For the sake of the discussion, we shall identify three traffic motifs representing a large fraction of RDMA workloads today. Unfortunately, those motifs also highlight RoCE’s shortcomings. Here, we focus on East-West (intra-) datacenter traffic as used in HPC, AI training and distributed inference, storage, as well as general microservice or Function as a Service (Faas) traffic.

\subsection{Incast (IN)}

An incast traffic pattern happens when multiple sources target the same destination process in a potentially uncoordinated but simultaneous traffic pattern. It is characterized by a number of source processes and a transaction size. It often appears stochastically in practice when a service is, by chance, requested by many uncoordinated clients at the same time. For example, imagine that 100 clients want to commit a 10 kiB write transaction to the same storage server. All clients may send at full bandwidth because they do not know about the upcoming congestion. The packets will quickly fill network buffers that can hinder other flows and eventually violate service level agreements (SLAs). The most challenging incast patterns are caused by transactions that are smaller than the bandwidth-delay product such that the congestion control mechanism cannot get a reliable signal before the transaction should be completed. We remark that growing bandwidths push more and more workloads into this critical region.

\subsection{Oblivious bulk synchronous (OBS)}

Many HPC and AI training workloads can be expressed in the oblivious bulk synchronous model (OBS) where computation steps are interleaved with global communication steps that often synchronize processes. \hl{Oblivious means that the communication pattern for an application depends on a small number of parameters (such as size or process count) and does not depend on the data that is processed. It can often be determined statically before the application is started.} For example, all collective operations in the Message Passing Interface (MPI) standard~\cite{mpi} are oblivious. Thus, OBS workloads can algorithmically avoid incast! The three-dimensional parallelism in deep learning training~\cite{hxmesh} is a typical example. OBS can be modeled by the number of processes, the duration of the computation, and the size of the communication (per endpoint). If both computation and communication are small, the overall workload is latency sensitive, a pattern that often appears in HPC and AI inference. Large communications that can often be found in AI training workloads are typically bandwidth-sensitive.

\subsection{Latency-sensitive (LS)}

For some workloads, message latency (and sometimes message rate) plays a central role. Some of those fall into the OBS category but others have complex, data-dependent, message chains that form critical performance paths in the application. Those are typically strong scaling workloads where the time to solution matters and inefficient execution must be tolerated. \hl{Large-scale simulations with strict deadlines such as weather forecasting and oil exploration fall into this category, but also some transaction processing or search/inference workloads.} Here, one has typically stringent (single-digit microsecond) latency requirements. 

\subsection{Deployment characteristics}

In addition to the traffic types, the deployment environment is also shifting. Newly emerging confidential compute ideas require all traffic to be encrypted on the wire. Ideally, traffic is encrypted and decrypted end-to-end in secure enclaves and no network equipment (NIC or switch) is to be trusted. Furthermore, and related, emerging multi-tenancy scenarios require managing tens of thousands of connections from a single host. Those are often supported by Smart NICs managing the resources such as bandwidth and security through rate limiting and filtering. Also, new, cost effective low-diameter and specialized topologies that require more advanced load balancing and routing become a necessity for extreme-bandwidth deployments~\cite{slimfly,hxmesh}. \hl{Many combinations of those requirements pose significant challenges on next-generation high-performance networks.}

\section{Where RoCE needs improvement}

Many of RoCE’s issues have been discussed in the past~\cite{rocky-road-for-roce} and many research works exist to propose various solutions~\cite{IRN}. \hl{Here, we outline potential improvements that we see and we relate them to the key workloads and deployment use-cases outlined above.} We now provide an itemized list of issues that could be improved for more efficient operation in Ethernet-based high-performance RDMA or Smart NIC systems. 

\subsection{1) PFC requires excessive buffering for lossless transport }

Priority Flow Control (PFC) lies at the very heart of converged Ethernet to enable lossless transport on each link. With PFC, the receiver monitors the available input buffer space. Once this buffer space falls below some threshold related to the bandwidth-delay product BW*RTT, it sends a PAUSE frame to the sender. At this time, BW*RTT/2 Bytes are already on the incoming wire but before the sender will receive the PAUSE frame, it will send another BW*RTT/2 Bytes. The minimal buffer requirement for fully lossless transfers would thus be BW*RTT + MTU\footnote{Maximum Transfer Unit}, where MTU is the maximum size of a packet. Yet, this would only support the case where packets are immediately drained at the receiver. Even the slightest delay in the forwarding may significantly reduce link utilization. 

The BW*RTT buffer space that covers the travel latency of the PAUSE message is often called “headroom buffer” and it is similar to the buffer required for credit-based flow control schemes such as those used in InfiniBand or Fibre Channel. In those, the receiver proactively sends credits (buffer allocations) to the sender keeping the input buffer space at an equilibrium, instead of reacting once it runs too full with PFC. Both schemes have their merits---a credit can travel proactively towards the source while a PFC scheme can be more reactive (late binding) when allocating shared buffer space to different source links. Both schemes need to essentially reserve BW*RTT space per link to just cover the round-trip control delay of the link, space that is lost for efficient forwarding.

\begin{figure}[h!]
	\centering
	\begin{subfigure}{1.0\columnwidth}
		\includegraphics[width=\textwidth]{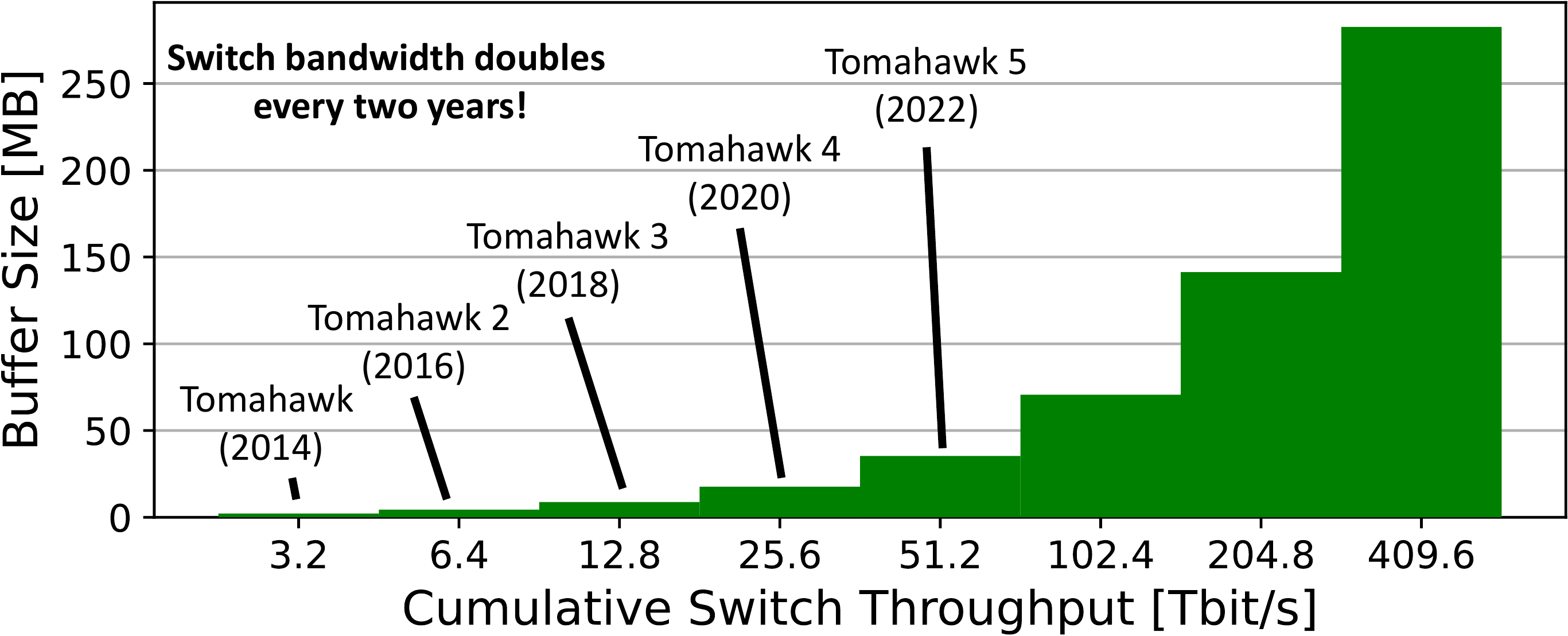}
		\caption{Intra-datacenter per-switch headroom buffer.}
		\label{fig:f1a}
	\end{subfigure}
	\hfill
	\begin{subfigure}{1.0\columnwidth}
		\includegraphics[width=\textwidth]{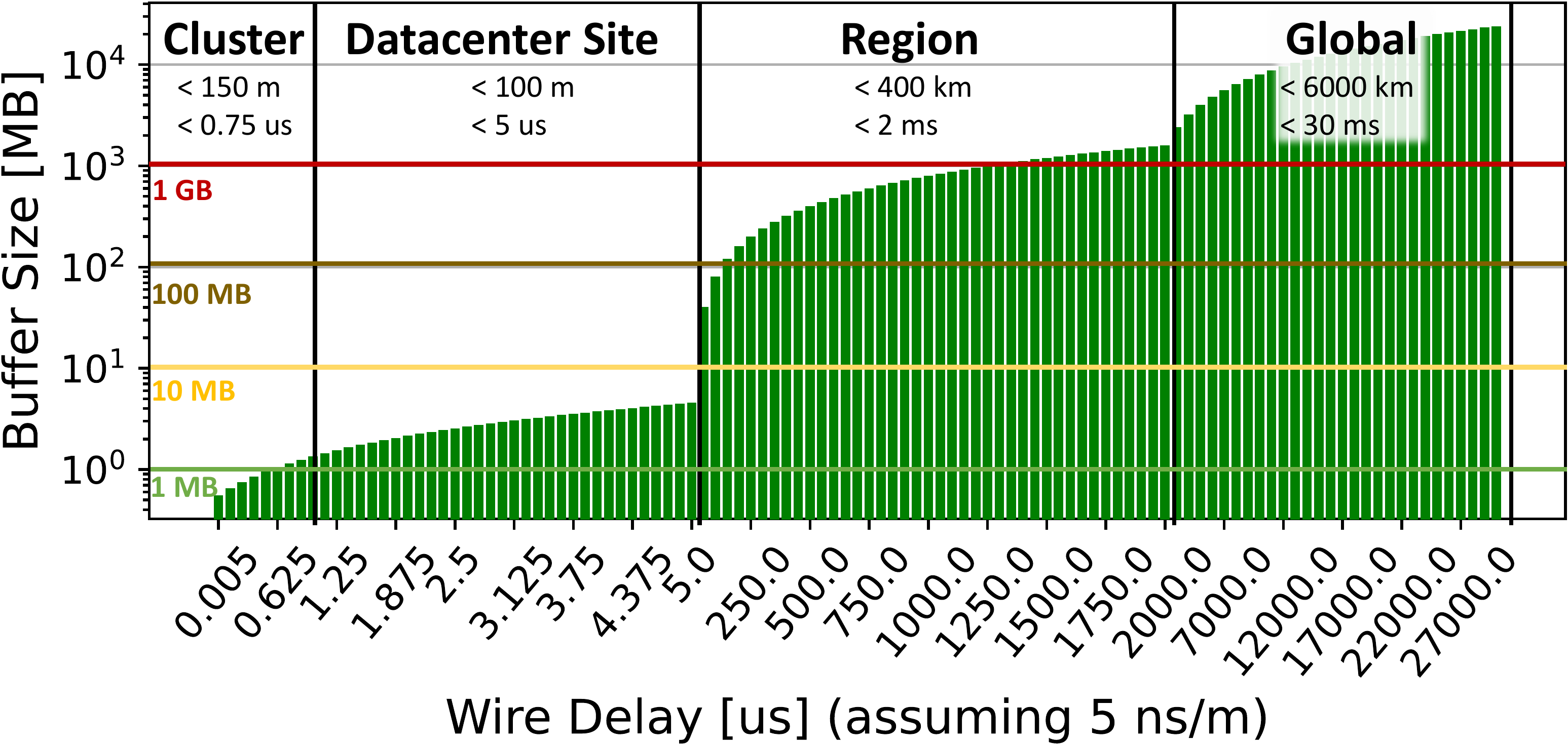}
		\caption{Varying distance per-port headroom buffer.}
		\label{fig:f1b}
	\end{subfigure}
	
	\caption{Headroom Buffer Requirements.}
	\label{fig:figures}
\end{figure}

In practice, buffer space is extremely valuable to ingest varying traffic peaks for temporal and spatial load balancing. Furthermore, just the required headroom buffer, that cannot be used for anything else without risking packet drops, puts a significant challenge for the scaling of next-generation switches. Figure~\ref{fig:f1a} shows the required headroom space (excluding other buffering!) for various switch generations assuming a 600 ns average latency (including arbitration, forward error correction (FEC), and wire delay) for 9 kB packets and 8 \hl{traffic priority classes} with separate buffers on a three-tier fat tree. Covering longer distances (and thus latencies) is also challenging as high-performance geo-replicated datacenters become common. Figure~\ref{fig:f1b} shows the needed per-port headroom buffer for the same configuration assuming 800G ports, a 5ns/m wire delay, and various deployment types. 

One may consider a lossy link-level protocol to repurpose these buffers for forwarding functions. Yet, this interacts with error handling protocols as we shall see soon. In any case, wasted buffer space is a general issue affecting all workloads that could benefit from the additional buffer if it was available for packet forwarding.

\subsection{2) Victim flows, congestion trees, PFC storms, and deadlocks}

Another issue stems from the fact that PFC stops a whole traffic class (encoded as only three bits) and all flows in it. This can lead to blocked victim flows: assume that we have two flows A and B sharing a link L. Flow A is not congested and could send at full bandwidth. However, flow B is blocked at some downstream port and fills up the input buffer of L. Eventually, L’s allocated buffer will be full with B’s packets and L sends a PAUSE frame. This frame also stops flow A, which could proceed independently---now, flow A is victimized by the PAUSE of flow B. Thus, flows that are not congested may be affected by other flows that are congested. This phenomenon is also known as Head of Line blocking.

Since any congestion of a downstream port will fill buffers upstream unless the endpoint congestion control protocol reacts, PFC events can quickly grow a ``congestion tree'' inversely following victimized flows in the network. Congestion trees are a general problem in lossless networks and are sometimes called PFC storms. It could be addressed by an even more fine-grained tracking of congestion, e.g., at the basis of individual flows instead of priorities. Yet, this requires the network switches to maintain flow state to identify individual flows~\cite{bpfc,sspaper}. One could also attempt to move congested flows into congested priorities dynamically, to avoid victims (cf. congestion isolation, P802.1Qcz). Another problem is that lossless lanes now consume already scarce traffic classes (separate buffer space). This takes an important resource from datacenter providers that already use such traffic classes for differentiated services such as elephant-flow backups, low-latency video conferencing, and others. Any traffic class used for RoCE (or other lossless) traffic is lost network-wide. 

Such congestion trees are particularly problematic for incast workloads where they can jam the whole network, especially in the context of packet-level adaptive or oblivious routing. Yet, the very low bandwidth per flow at the incast link means that, in theory, these flows would need very little network buffering to saturate the link. The purely rate-based nature of RoCE’s congestion control allows sources to inject (too) many packets that quickly fill network buffers. For example, a window-based scheme would allow the administrators to directly control the network-wide buffer occupancy of each flow. 

Any lossless scheme with limited buffering suffers from deadlocks if the routing allows for cycles to form. This can be avoided with cycle-free routing schemes or special buffering strategies---both come at a (small) cost. Even if routes are generally deadlock free, transient states occurring after link failures can lead to deadlocks. Avoiding those is harder, however, one can configure packet timeouts in switches to resolve this problem dynamically. 

\subsection{3) Go-back-N retransmission}

RoCE was designed for very simple hardware following InfiniBand’s in-order and credit-based lossless transport. This implies that packets can only be dropped if they are corrupted by bit errors, a very rare event. Thus, retransmission logic can be simple: if the receiver detects a gap in the packet stream (i.e., a skipped sequence number), it sends a negative acknowledgement (NACK) to the sender and drops all later packets. The sender then retransmits all packets beginning with the lost one. This scheme essentially discards and retransmits a full end-to-end BW*RTT (bandwidth-delay product) worth of data. 

Let us assume a three-tier fat tree network with 800 Gb/s link speed and a worst-case per-hop latency of 600 ns. The total RTT as observed by an endpoint would be 3.6 us\footnote{\hl{we roughly approximate end-to-end latency as six  hops}}. The effective bit error rate on each link can be as high as 1e-12 (as proposed by the Ethernet specification~\cite{802/44}) and we assume 9 kiB frames, the probability of losing a single frame is 3.3e-8 (see Appendix \ref{appendix:frame-loss-prob} for derivation). Thus, the total expected bandwidth loss due to go-back-n would be a negligible 0.00013\%.

A bigger issue with the simple go-back-n scheme is that it does not support multi-pathing or out-of-order delivery. Any two packets passing would trigger an expensive retransmission event losing a full BW*RTT transmission. Latest generations of RoCE NICs introduce selective retransmission to mitigate this problem. Yet, those are often limited. For example NVIDIA’s ConnectX-6 adapter does not support adaptive routing of tag matching with selective retransmission enabled.\footnote{ConnectX-6 DX firmware release notes v22.27.1016} %
Go-back-n has one interesting advantage though: if a bit error happens and the packet is dropped (silently) by the lower layers, the error is detected immediately once the next packet arrives. Other schemes that support out-of-order delivery would need to wait for a timeout to expire at the sender, potentially leading to much higher recovery times and jitter. Thus, when designing new transport protocols, one needs to consider all these tradeoffs carefully!

\subsection{4) Congestion control and colocation with other traffic}

RoCE’s default congestion control relies on a very simple rate control that is intimately linked to the lossless transport assumption. Many researchers have recognized that this simple mechanism does not integrate well with other traffic such as TCP/IP and generally can be improved in the datacenter environment. Mechanisms such as DCQCN~\cite{dcqcn}, TIMELY ~\cite{timely}, and HPCC~\cite{hpcc} build on RoCE to improve the transport of flows. Most RoCE deployments today use non-standard congestion control mechanisms which makes interoperability between vendors, or even different hardware generations of the same vendor, hard. This is due to the fact that congestion control remains a tough problem and it is likely that different workloads require different tuned versions of the protocol.

For example, the typically repetitive endpoint-congestion-free bulk data transfers in oblivious synchronous workloads could quickly be learned or even be statically configured based on the expected traffic pattern~\cite{hxmesh,RoCECCDNN}. Highly-dynamic incast scenarios require coordinating multiple senders either through the receiver or network signals. Latency-sensitive workloads with small messages that are smaller than the bandwidth-delay product can be most problematic, especially if they appear in an unpredictable data-driven communication pattern. Those may need to rely on switch buffering to ingest temporary load-imbalance at the network level. In general, congestion control schemes are and will remain a research focus with constant tuning even after deployment. Co-existing with different traffic types such as TCP or QUIC will also require constant adoption. Thus, such schemes should not only be fast and cheap in hardware but also be flexible and support a wide range of parametrizations. 

Another line of argument considers switch queue size and occupancy. Datacenter switches traditionally have large (deep) buffers to accommodate traffic bursts without dropping to accommodate the slow end-to-end rate adjustment. On the other hand, switches used in HPC usually operate lossless with very shallow buffers and stiff back-pressure due to their reliable link-level flow control mechanisms~\cite{sspaper}. Also, HPC network topologies have usually lower diameter than datacenter deployments~\cite{diam2}. Thus, HPC deployments support lower-latency operations because small packets are less likely to wait in buffers behind longer flows. Datacenter networks with RoCE are often combining both inefficiently: they use a lossless transport with all its issues with relatively large-buffered switches. Many modern congestion control mechanisms thus aim at keeping the buffer occupancy generally low, leaving this very expensive resource unused!

\subsection{5) Header sizes, packet rates, scalability}

RoCEv2 uses full Ethernet L2 and UDP/IP headers in addition to InfiniBand’s Base Transport Header (BTH). Thus, the header overhead per packet is substantial: 22 Bytes L2, 20 Bytes IP, 8 Bytes UDP, and 12 Bytes BTH and 4 Bytes ICRC make a total of 66 Bytes per packet. Locally-routed InfiniBand, for example, has only a total header size of 20 Bytes: 8 Bytes for the Local Routing Header, and 12 Bytes for the BTH. Other HPC protocols have headers with less than 40 Bytes. 

This impacts both the raw packet rate as well as processing overhead and cost as more complex headers require more header processing. Just the packet rate for small payloads could be problematic. Let us assume 8 Byte messages as an example for a single-element reduction operation for conjugate gradient solvers or fine-grained global graph updates. The maximum rate (without headers) on an 800 Gb/s link would be 12.5 \hl{Giga-packets per second (Gpps)}. With IB headers, that rate would decrease to 3.5 Gpps and with RoCEv2 headers to 1.4 Gpps. The packet would be nearly 90\% header overhead! And we are ignoring additional protocol headers for MPI or RDMA endoints. Yet, given that NIC packet processing is currently slower ($<$1 Gpps per NIC), the header size may not be the biggest issue. Furthermore, NICs need to process acknowledgment packets, which could be especially challenging for selective acknowledgment and retransmission protocols. The high user-level and protocol message rates require parallel processing in the NIC given the mostly stagnant clock rates. 

RoCE’s packet format is closely linked to InfiniBand’s verbs which has connections between queue pairs (QPs) as its basic concept. The size of the context state for a single connection depends on the implementation details but large-cluster all-to-all connectivity may be problematic. Each queue pair at least needs to keep connection information and state such as sequence number and destination address and queue pair number. Connection state can be relatively large, up to 1 kB per connection in some implementations.

Small packets are often important in latency-sensitive workloads, some of which are bound by the rate at which the NIC can issue new messages. Slimmer headers would potentially decrease latencies and increase message rates while allowing for a more efficient bandwidth utilization.

\subsection{6) No support for smart stacks}

As network overheads become more important in datacenter workloads, more intelligent stacks are designed. For example, the QUIC protocol allows to push transport processing to the application which can define application-specific protocols. This enables running different protocols for different service requirements, such as latency-insensitive video streaming, latency sensitive audio-conferencing, or generally resilient but large backup traffic. \hl{RoCE’s philosophy of hardware acceleration does not support different transport protocols, even if the user-level stack would be able to specify additional properties of the traffic (e.g., mark messages as resilient to out-of-order delivery).}

Emerging Smart NICs lead to new opportunities in this area where user-configurable kernels could perform packet and protocol processing on the NIC~\cite{spin}. Additionally, in-network telemetry (INT) can provide additional signals for these protocols to react accordingly. Thus, even if the stack has additional knowledge about the traffic types, today’s RoCE forces it into a relatively simple and inflexible protocol that cannot take full advantage of this knowledge.

\subsection{7) Security}

RoCE is known to have several security issues~\cite{sRDMA,Redmark}, especially in multi-tenant contexts. Many of those issues stem from the fact that protocol security, authentication, and encryption have played a minor role at the design time. Yet, today, such properties are much more important. 

IPSEC can be used to protect L3 headers and payload but would need to be enabled on a per-queue-pair basis such that no two tenants share a set of keys. This can be quite costly in terms of connection context overhead and performance. Furthermore, RoCE does not support sub-delegation of memory regions to other nodes. Both issues can be addressed with modern key-derivation protocols~\cite{sRDMA}.

\subsection{8) Link-level reliability}

The move towards higher transceiver speeds leads to more complex encoding and modulation schemes running at growing frequencies. With 50G lanes, Ethernet moved from the simple two-voltage level NRZ to four-voltage level PAM4 encoding. Today’s 100G lanes run at 25 GHz, requiring the receiver to distinguish four levels within a fraction of a nanosecond. The signal degradation in cables and connectors as well as the increasingly complex analog circuitry lead to higher bit-error rates going to a bit-error rate (BER) as high as 1e-4 soon. 

Forward-error correction (FEC) has been introduced to avoid excessive end-to-end retransmissions due to dropping of corrupted packets in the network. Ethernet aims at a 1e-12 BER at the link level and currently employs a Reed-Solomon code on 10-bit symbols using a block of 514 such symbols with 30 additional encoding symbols (RS544). This enables the receiver to correct 15 random bit errors and up to 150 consecutive (burst) bit errors. Other FEC codes such as LL-FEC (RS272, half size as RS544) and Firecode provide lower latency but also lower protection against bit errors. 

\begin{figure}[h!]
	\centering
		\includegraphics[width=\columnwidth]{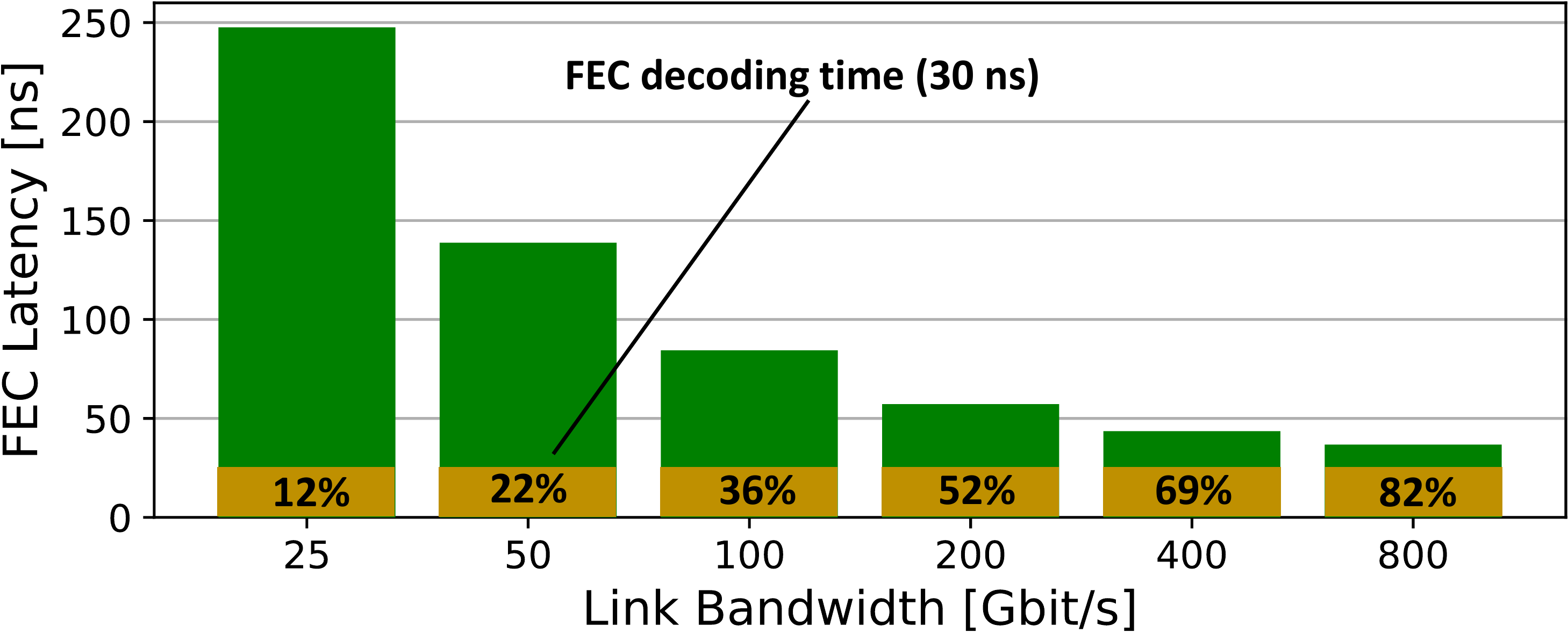}
	\caption{RS544 FEC latency breakdown.}
	\label{fig:f2}
\end{figure}

Generally, FEC comes at a latency and energy cost that falls into two categories: (1) accumulating the 5,140 bits of data and (2) encoding and decoding the code symbols. The former decreases with the link bandwidth and the latter depends on the implementation, varying from 20 to 100 ns in practice. Figure~\ref{fig:f2} shows the projected RS544 FEC for different link bandwidths. 

For a constant RS544 FEC, the latency reduces for faster link bandwidths but will not go below the FEC computation overhead. However, faster lanes may lead to significantly higher bit error rates. In fact, RS544 may not be able to correct the projected 1e-4 BER to the desired 1e-12. Thus, future Ethernet standards may move to more complex FEC mechanisms that may increase the latency significantly.

An alternative approach is used in PCIe, which also deals with relatively high BER due to complex connectors but is designed as a low-latency local interconnect targeting around 5 ns. For example, the upcoming PCIe 6.0 specification protects a block of 242 Bytes with 6 Bytes of FEC together with an additional 8 Byte CRC. The receiver first uses the FEC to correct some bit errors and then checks the CRC. If this check fails, it initiates a simple link-layer retransmission protocol to request the data again. The FEC reduces the bit error rate from 1e-4 to 1e-6 and the CRC then triggers retransmission with probability of less than 1e-5. The latency addition due to FEC is less than 2ns and the bandwidth reduction due to retransmission less than 2\%. %
The challenge for Ethernet are longer links leading to higher link-latencies.

\section{System issues}

\begin{figure}[h!]
	\centering
	\includegraphics[width=\columnwidth]{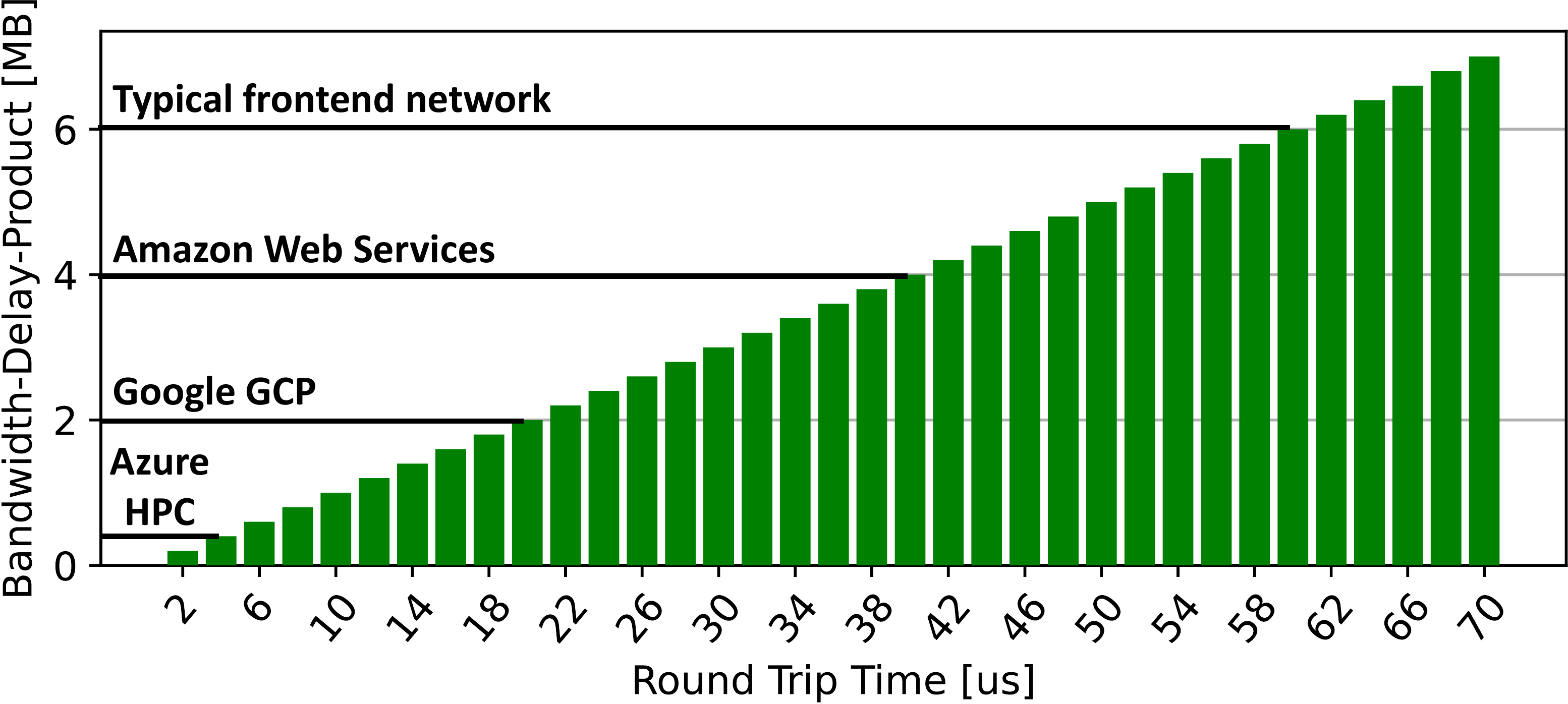}
	\caption{Bandwidth-delay-product vs. Round-Trip-Time (numbers from De Sensi et al.~\cite{cloudnoise})).}
	\label{fig:f3}
\end{figure}

Growing link-level and thus end-to-end latencies can lead to more issues at the system level. Higher latencies lead to higher buffer occupation and energy consumption. Less obviously, higher latencies lead to less efficient congestion control: messages that are transmitted faster than a single RTT cannot benefit from congestion control mechanisms that rely on receiver-based notifications. The bad case of incast with small messages thus gets worse or at least more common because the size of a ``small message'' increases. Figure~\ref{fig:f3} shows the size of the bandwidth delay product for some realistic latencies in datacenters today showing that even 1 MiB messages can be considered ``too small'' for effective incast handling by throttling the sender. Thus, problematic incast patterns may become more common with higher latencies!

In other words, if a system can throttle the sender fast enough, it can reduce the message size below which incast is a problem. This can be achieved by lowering latencies or having  switches report incast congestion directly to the source (without bouncing through the receiver). Furthermore, if only very small messages create bad-case incasts, switch buffers may simply ingest them in the common case without even running out of resources. This may be amplified along incast trees where multiple sets of switch buffers can ingest transient incast messages, of course, potentially leading to congestion trees in the network. Such whole-systems issues remain an open discussion but it seems that lower latency generally simplifies them.

One also needs to pay attention to other aspects of the overall stack that can be quite complex. For example, simple and clear (remote) memory semantics are tricky to define, reason about, and implement correctly~\cite{featherweight-rdma}. Furthermore, the fact that process-local virtual addresses are exposed to remote hosts can be problematic for security and performance. One could think of a scheme with addressing relative to a memory region~\cite{Portals4}. From a security perspective both schemes have their weaknesses: exposing addresses allows learning about the remote process, yet fixed offsets are much simpler to guess for an attacker~\cite{Redmark}. We note that these are general problems for all RDMA systems and not specific to RoCE.

Routing and load balancing remains an open challenge---most HPC networks use packet-level adaptive routing with relatively advanced in-network mechanisms~\cite{sspaper} while most datacenter networks use simple oblivious ECMP driven by the endpoints that change header fields to guide path selection in very simple ways. The granularity of such ECMP load balancing in data centers ranges from traditionally full flows to recently considered flowlets. Flowlets are consecutive sequences of packets that have a sufficient gap between them that flowlets cannot pass each other even when sent along different routes. Such gaps can be introduced by delaying packets or appear naturally. More recently, datacenter networks are looking towards more fine-grained mechanisms for load balancing. Another challenge is the requirement of some applications that messages be delivered in order. In general, out-of-order granularity and capabilities depend heavily on application requirements and the capabilities of the endpoint NICs. Finer and more out-of-order capabilities simplify network load balancing. 

\section{Predictions}

\hl{Based on all these points, we predict that academia and industry will revisit datacenter Ethernet. This next-generation Ethernet will likely support lossy and lossless transport modes for RDMA connections to allow intelligent switch-buffer management.} This will make the provisioning of headroom buffer optional and avoid the other problems such as victim flows and congestion trees of lossless networking. Next-generation Ethernet is also unlikely to adopt go-back-n retransmission semantics but opt for more fine-grained mechanisms such as selective acknowledgments. Furthermore, it will likely make congestion management part of the specification. Special attention will be paid to colocation with other flows, espcially in lossy traffic classes. The protocols will be designed in a flexible way to support smart networking stacks and security will finally become a first-class citizen. We may also see innovations in headers and reliability approaches as well. 

Such modernizations will drive a new high-performance networking ecosystem for AI, HPC, and storage systems that are at the heart of hyperscale datacenters. This development will conclude the convergence of HPC and datacenter networks!

\bibliographystyle{ieeetr}
\bibstyle{ieeetr}
\bibliography{ethernet_rdma_problems}

\begin{IEEEbiography}{Torsten Hoefler}{\,}
is a Professor of Computer Science at ETH Z\"urich. His research interests revolve around large-scale high-performance systems and networks for HPC and AI. He is a fellow of the ACM and IEEE as well as a member of Academia Europaea. For more information visit  \url{http://htor.ethz.ch}.
\end{IEEEbiography}

\begin{IEEEbiography}{Duncan Roweth}{\,}
is a Senior Distinguished Technologist in the Slingshot Business Unit CTO office at HPE.   He joined HPE in Jan 2020 with the acquisition of Cray.  While at Cray he worked on three generations of HPC network.  He has been in a leading figure in the Slingshot program since its inception. Duncan holds a Ph.D. from the University of Edinburgh.  
\end{IEEEbiography}

\begin{IEEEbiography}{Keith Underwood}{\,}
is a Senior Distinguished Technologist at Hewlett Packard Enterprise. He is an architect in the Slingshot program focusing on network interface architecture.  
\end{IEEEbiography}

\begin{IEEEbiography}{Robert Alverson}{\,}
is a Distinguished Technologist at Hewlett Packard Enterprise. Bob has a Master of Science degree from Stanford University in Electrical Engineering. He has a long history in HPC interconnects and was a lead architect of Slingshot high speed network interconnect, which combines HPC performance with Ethernet compatibility in a dragonfly network.
\end{IEEEbiography}

\begin{IEEEbiography}{Mark Griswold}{\,}
is a Distinguished Engineer and Switch Architect at Broadcom.
His primary research interests are high-performance interconnects, computer
architecture, device architecture and workloads.  He obtained a B.S. in
Mathematics \& Computer Science and a B.S. in Computer Engineering, both from
Carnegie Mellon University.
\end{IEEEbiography}

\begin{IEEEbiography}{Vahid Tabatabaee}{\,}
is a Distinguished Engineer and Switch Architect at
Broadcom. He received his Ph.D. from the University of Maryland, College Park.
His research interests are in algorithm design and performance analysis for congestion control
and traffic management in networks. His email is vahid.tabatabaee@broadcom.com.
\end{IEEEbiography}

\begin{IEEEbiography}{Mohan Kalkunte}{\,}
is the Vice President of Architecture and Technology responsible for the architecture development of switches for Enterprise, Data Center, and Service Provider markets at Broadcom.  He is an IEEE fellow and has over 150 patents. His email is mohan.kalkunte@broadcom.com.
\end{IEEEbiography}

\begin{IEEEbiography}{Surendra Anubolu}{\,}
is a Distinguished Engineer at Broadcom in the Switch Group. He is currently working on benchmarking and enhancing performance of distributed AI work loads and
telemetry for network applications. He holds an MS
from Indian Institute of Science, Bangalore.
\end{IEEEbiography}

\begin{IEEEbiography}{Siyuan Shen}{\,}
received his MEng degree in Computing from Imperial College London and is currently a Ph.D. student in the Scalable Parallel Computing Lab at ETH Zurich. His primary research interests include distributed computing, networking, and distributed machine learning.
\end{IEEEbiography}

\begin{IEEEbiography}{Moray McLaren}{\,}
is a Principal Engineer at Google.   His research interests include
future data center networking, and interconnects for High Performance Computing and
Machine Learning.  His e-mail is moray@google.com.	
\end{IEEEbiography}

\begin{IEEEbiography}{Abdul Kabbani}{\,}
is a Principal Network HW Architect at Microsoft. His research interests include congestion management algorithms in high-performance and frontend networking. He received his PhD from Stanford University.
\end{IEEEbiography}

\begin{IEEEbiography}{Steve Scott}{\,}
is a Technical Fellow and Corporate Vice President of Azure Hardware Architecture at Microsoft. His research interests are in high performance system and network architecture.  He is an IEEE and ACM Fellow, and received his PhD from the University of Wisconsin at Madison.
\end{IEEEbiography}

\appendix
\subsection{Derivation of Frame Loss Probability}
\label{appendix:frame-loss-prob}

\vspace{.5em}
To measure the performance of a specific $RS(n, k)$ scheme under the assumption that a random-error model is used, we need to find the probability of losing an Ethernet frame given a pre-defined input bit error rate ($BER_{in}$) and the number of hops in the network. As a first step, we can calculate the input symbol error rate ($SER_{in}$), i.e. the probability of an FEC symbol being corrupted, as:
\begin{align*}
SER_{in} &= 1 - (1 - BER_{in})^m\\
    &\approx m \cdot BER_{in} \text{ (for small $BER_{in}$)}
\end{align*}
where $m$ is the number of bits in an FEC symbol. The sub-expression $(1 - BER_{in})^m$ represents the probability of having no errors in a symbol.

The number of symbols in an FEC codeword that can be corrected by a $RS(n, k)$ scheme is expressed as $t = \lfloor \frac{n - k}{2} \rfloor$, which signifies that after decoding, the uncorrectable codeword error rate ($CER$) is:
\[CER = \sum_{i = t + 1}^{n} \underbrace{\binom{n}{i} SER_{in}^{i} (1 - SER_{in})^{n - i}}_{\substack{\text{Probability of having exactly $i$} \\ \text{corrupted symbols in a codeword}}} \]

Since an Ethernet frame can only be properly received when all of its constituent codewords are correctable, we can compute the frame error rate ($FER$) as:
\begin{align*}
FER &= 1 - (1 - CER)^{1 + \lfloor \frac{\text{frame size}}{\text{codeword size}} \rfloor}\\
    &\approx \Bigl(1 + \lfloor \frac{\text{frame size}}{\text{codeword size}} \rfloor \Bigr) \cdot CER
\end{align*}
where the average number of codewords that a frame occupies is denoted by $1 + \lfloor \frac{\text{frame size}}{\text{codeword size}} \rfloor$.

After obtaining the $FER$ per link, the frame loss probability $P$ is simply:
\begin{align*}
P &= 1 - (1 - FER)^{hops + 1}\\
  &\approx (hops + 1) \cdot FER
\end{align*}

\end{document}